\documentclass{sigchi}

\usepackage{balance}  
\usepackage{graphics} 
\usepackage{times}    
\usepackage{url}      

\usepackage{fancyvrb}
\usepackage{wrapfig}

\usepackage[T1]{fontenc}
\usepackage{inconsolata}

\def\levelThree#1{\subsubsection{#1}}

\newcommand{\ie}{{\emph{i.e.}}}
\newcommand{\eg}{{\emph{e.g.}}}

\newcommand{\mdash}{\hspace{0.02in}---\hspace{0.02in}}

\newcommand{\sns}{\ensuremath{\textsc{Sketch-n-Sketch}}}
\newcommand{\little}{\texttt{little}}

\newcommand{\powerpoint}{{PowerPoint}}
\newcommand{\illustrator}{{Illustrator}}
\newcommand{\keynote}{{Keynote}}

\newcommand{\bez}{B\'{e}zier}

\newcommand{\myurl}[1]{\url{#1}}

\newcommand{\refSecOverview}{the Overview section}
\newcommand{\refSecDraw}{the Tools for Drawing Shapes section}
\newcommand{\refSecRelate}{the Tools for Relating Features section}
\newcommand{\refSecGroup}{the Tools for Grouping Shapes section}
\newcommand{\refSecEvaluation}{the Evaluation section}

\newcommand{\codeSize}{\small}

\newcommand{\miniSepOne}{\hspace{0.01in}}

\newcommand{\miniSepThree}{\hspace{0.03in}}

\newcommand{\ttlparen}{\ensuremath{\texttt{(}}}
\newcommand{\ttrparen}{\ensuremath{\texttt{)}}}

\newcommand{\num}[1]{\ensuremath{\texttt{#1}}}
\newcommand{\op}[1]{\ensuremath{\texttt{#1}}}

\newcommand{\parens}[1]{\ensuremath{\ttlparen{#1}\ttrparen}}

\newcommand{\expApp}[2]{\ensuremath{\parens{{#1}\miniSepThree{ }{#2}}}}
\newcommand{\expAppTwo}[3]{\ensuremath{\expApp{#1}{\twoThings{#2}{#3}}}}

\newcommand{\expStr}[1]
  {\ensuremath{\texttt{'{#1}'}}}

\newcommand{\labeledNum}[2]{\ensuremath{{#1}^{#2}}}

\newcommand{\ttNumNameVerb}[2]{\labeledNum{\num{#1}}{\num{#2}}}

\newcommand{\twoThings}[2]{\ensuremath{{#1}\miniSepThree{#2}}}

\newcommand{\figSyntaxEnd}{\end{array}$}

\newcommand{\figSyntaxBegin}{$\begin{array}{rcl}}

\newcommand{\aSubst}[2]{{#1}\mapsto{#2}}

\newcommand{\helperop}[1]{\ensuremath{\mathsf{#1}}}

\newcommand{\myEquation}[2]{\ensuremath{{#1}={#2}}}

\newcommand{\solveOne}[4]{\ensuremath{\helperop{SolveOne}(#1,#2,\myEquation{#3}{#4})}}

\newcommand{\solveAndUpdate}[4]
  {(\aSubst{#2}{\solveOne{#1}{#2}{#3}{#4}})}

\newcommand\verbBoxBlue[1]
  {\textcolor[rgb]{0.39,0.58,0.93}{\miniSepOne{\miniSepOne{\textbf{#1}}\miniSepOne}}\miniSepOne}

\newcommand\verbOrange[1]
  {\textcolor[rgb]{0.84,0.40,0.00}{\textbf{#1}}}

\makeatletter
\def\url@leostyle{%
  \@ifundefined{selectfont}{\def\UrlFont{\sf}}{\def\UrlFont{\small\bf\ttfamily}}}
\makeatother
\def\pprw{8.5in}
\def\pprh{11in}

\usepackage[pdftex]{hyperref}
\hypersetup{
pdftitle={Semi-Automated SVG Programming via Direct Manipulation},
pdfauthor={Brian Hempel and Ravi Chugh},
pdfkeywords={UIST 2016, SIGCHI},
bookmarksnumbered,
pdfstartview={FitH},
colorlinks,
citecolor=black,
filecolor=black,
linkcolor=black,
urlcolor=black,
breaklinks=true,
}

\begin{document}

\title{Semi-Automated SVG Programming via Direct Manipulation}

\numberofauthors{1}
\author{
  \alignauthor Brian Hempel \hspace{0.35in} Ravi Chugh\\[3pt]
    \affaddr{Department of Computer Science, University of Chicago, USA}\\
    \texttt{\{brianhempel,rchugh\}\hspace{0.02in}@\hspace{0.02in}uchicago.edu}
}


\maketitle

\begin{abstract}
Direct manipulation interfaces provide intuitive and interactive
features to a broad range of users, but they often exhibit two
limitations: the built-in features cannot possibly cover all
use cases, and the internal representation of the content
is not readily exposed.
We believe that if direct manipulation interfaces were to
(a) use general-purpose programs as the representation format, and
(b) expose those programs to the user,
then experts could customize these systems in powerful 
new ways and non-experts could enjoy some of the benefits of
programmable systems.

In recent work, we presented a prototype SVG editor called \sns{}
that offered a step towards this vision. In that system, the user
wrote a program in a general-purpose lambda-calculus to generate a
graphic design and could then directly manipulate the output to indirectly
change design parameters (\ie{} constant literals) in the program in
real-time during the manipulation. Unfortunately, the burden of
programming the desired relationships rested entirely on the user.

In this paper, we design and implement new features for \sns{} that
assist in the programming process itself.
Like typical direct manipulation systems, our extended \sns{} now provides
GUI-based tools for drawing shapes, relating shapes to each other, and grouping shapes together.
Unlike typical systems, however, each tool carries out the user's intention by transforming their general-purpose program.
This novel, semi-automated programming workflow allows the
user to rapidly create high-level, reusable abstractions in the program while at the
same time retaining direct manipulation capabilities.
In future work, our approach may be extended with more graphic design features or realized for other application domains.
\end{abstract}


\category{H.5.2.}{Information Interfaces and Presentation (e.g. HCI)}{User Interfaces}
\category{D.2.6.}{Software Engineering}{Programming Environments}
\category{D.3.3.}{Programming Languages}{Language Constructs and Features}

\section{Introduction}

Direct manipulation interfaces~\cite{Shneiderman1983} provide
a broad range of users the tools to author content in a variety of
application domains. In addition to intuitive and immediate
feedback, full-featured direct manipulation systems
(\eg{} Adobe \illustrator{},
Microsoft \powerpoint{}, and
Apple \keynote{})
provide scores of built-in features
such as rulers, snap-to alignment, grouping, and animations
for operations common to a target domain.

Nevertheless, experts and novices alike must often resort to
mundane, repetitive tasks\mdash{}such as excessive copy-and-pasting
and secondary edits to keep conceptually related objects
in sync\mdash{}that could be avoided with general-purpose programming.
Some direct manipulation systems provide APIs for
customization, but they are typically not connected to the main
application in lightweight, easy-to-use ways.

In response,
researchers have recently proposed several approaches that attempt
to strike a balance between intuitive interactivity
and expressive programmability.
We believe these prior efforts can be classified into two broad
categories.

\levelThree{Adding Programming to Direct Manipulation}

One approach has been to extend a mostly traditional direct manipulation
system with ``structured'' programming features, by which we mean
design choices that restrict the expressiveness of the language
(\eg{} a domain-specific language) or provide limited code editing tools
(\eg{} a block editor).
Software tools in this category, such as
Drawing Dynamic Visualizations~\cite{VictorDrawing},
Apparatus~\cite{Apparatus}, and Programming by Manipulation~\cite{PBM},
tend to favor direct manipulation and rely on programming as
a last-resort scenario.

\levelThree{Adding Direct Manipulation to Programming}

In contrast, another approach has been to extend a mostly traditional,
general-purpose programming language with direct manipulation features.
Software tools in this category, such as Live Programming~\cite{WangEtAl,McDirmid,LivePBE}
and our previous version of \sns{}~\cite{sns},
expect that users will work closely with
text-based programs but also provide ways for directly manipulating
output values to indirectly edit the program.

The approaches in the first category have significant
merit, especially for specific application domains.
However, we favor the second approach\mdash{}that
general-purpose languages can be the foundation upon which
to build full-featured user interfaces\mdash{}because of the potential
for new techniques to augment programming methodologies in general.
Unfortunately, our previous effort required users to carry out
the lion's share of the design work programmatically; only then could the user
directly manipulate the output to indirectly modify design parameters
(\ie{} constants) in the program~\cite{sns}.

\subsection{Our Approach and Contributions}

We present new
direct manipulation features for the \sns{} Scalable Vector Graphics
(SVG) editor that interactively assists the user in building a program
in a high-level, general-purpose programming language.

Specifically, our paper makes the following contributions:

\begin{itemize}

\item

We design novel direct manipulation features for
(1) drawing new shapes, (2) relating features among shapes, and
(3) building reusable abstractions from groups of shapes.
Each of the tools in our Draw, Relate, and Group workflow is
paired with a program transformation that automates some of the
mundane parts of programming, significantly reducing
the programming burden.

\item

We implement our design in the context of the \sns{} system.
Our implementation is open-source and publicly available on the Web.

\item

We demonstrate how our new \sns{} implementation can be used to author
several graphic designs where the bulk of the programming is done
automatically by the system. Videos of some of these examples are available on
the Web.\footnote{\ \url{http://ravichugh.github.io/sketch-n-sketch}}

\end{itemize}

As a result, the user now starts by drawing and interacting with
shapes on the canvas and receives guidance from the system in turning the
desired relationships into a high-level, readable program, which then
enables the interaction capabilities developed in our previous
work for manipulating design parameters
via direct manipulation.

We believe that our ideas serve as another step towards the long-term
goal of truly achieving a harmonious combination of programming and direct
manipulation.

\section{Overview}
\label{sec:overview}

We now provide an overview of how our extensions to \sns{} interactively
help the user build a program that implements a reusable graphic
design. The workflow in \sns{} can be viewed as a series
of three phases
(Drawing Shapes, Relating Features, and Grouping Shapes)
which, in practice, may overlap and be interleaved.
In this section, we will explain each phase in terms of a running
example. Then, in three subsequent sections, we discuss each phase
in more detail.

\setlength{\intextsep}{6pt}%
\setlength{\columnsep}{10pt}%
\begin{wrapfigure}{l}{0pt}
\includegraphics[scale=0.25]{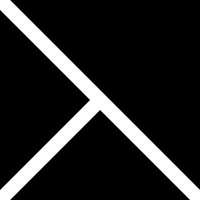}
\end{wrapfigure}
Our overview example is to build the \sns{} logo, shown on the
left, which consists of three triangular positive areas
separated by two equal-width negative stripes.
We would like to implement this logo in a way that makes the design
parameters (colors, width, and size) easy to change.
Starting from an initially empty program that does not draw any
shapes, we will be able to create the final program entirely using the
direct manipulation tools in \sns{}. In \refSecEvaluation{}, we
will describe examples that require some edits to the source code.

\subsection{Drawing Shapes}

The programming language in \sns{}, called \little{}, is a
general-purpose, untyped lambda-calculus, and the values produced
by a \little{} program are translated to the SVG format
for rendering~\cite{sns}.
We extend \sns{} with direct manipulation drawing tools for several
common shapes, so that the user does not have to write \little{} code
to add new shapes.
When a shape is drawn in the canvas, the editor adds
a corresponding definition to the \little{} program such that,
when re-evaluated, the program produces the new shape.

\begin{wrapfigure}{r}{0pt}
\includegraphics[scale=0.25]{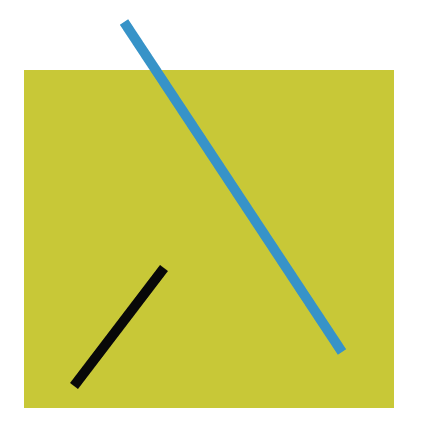}
\end{wrapfigure}
In order to implement the logo,
our first step is to pick an underlying representation for
the design. One option is to use a single rectangle
in the background and two lines in the foreground.
When we draw a rectangle and two lines very roughly positioned
on top of it (as shown on the right),
\sns{} generates the program in \autoref{fig:sns-logo-v1}.

\begin{figure}[t]
\begin{Verbatim}[commandchars=\\\{\},
                 codes={\catcode`\$=3\catcode`\^=7\catcode`\_=8},
                 numbers=left,numbersep=-6pt,
                 fontsize=\codeSize]
   (def rect1
     (let [left top right bot] [31 100 216 269]
     (let bounds [left top right bot]
     (let color 60
       [ (rectangle color 'black' '0' 0 bounds) ]))))
   
   (def line2
     (let [x1 y1 x2 y2] [81 76 190 241]
     (let [color width] [202 5]
       [ (line color width x1 y1 x2 y2) ])))
   
   (def line3
     (let [x1 y1 x2 y2] [56 258 101 199]
     (let [color width] [383 5]
       [ (line color width x1 y1 x2 y2) ])))
 
   (blobs [ rect1 line2 line3 ])     ; "Main" Expression
\end{Verbatim}
\vspace{0.05in}

\caption{Overview example after drawing new shapes.}
\label{fig:sns-logo-v1}
\end{figure}

The program structure comprises a series of top-level definitions
followed by a ``main'' expression that defines the output SVG canvas.
Notice that the formatting and identifier names are
designed to be easy for the user to read and edit.
The library functions \verb+rectangle+ and \verb+line+ draw the
corresponding SVG primitives, and the \verb+blobs+ function denotes
list concatenation.
We will describe the program in \autoref{fig:sns-logo-v1} in more
detail in \refSecDraw{}.

\levelThree{Live Synchronization: Direct Manipulation and Widgets}

\begin{wrapfigure}{r}{0pt}
\includegraphics[scale=0.25]{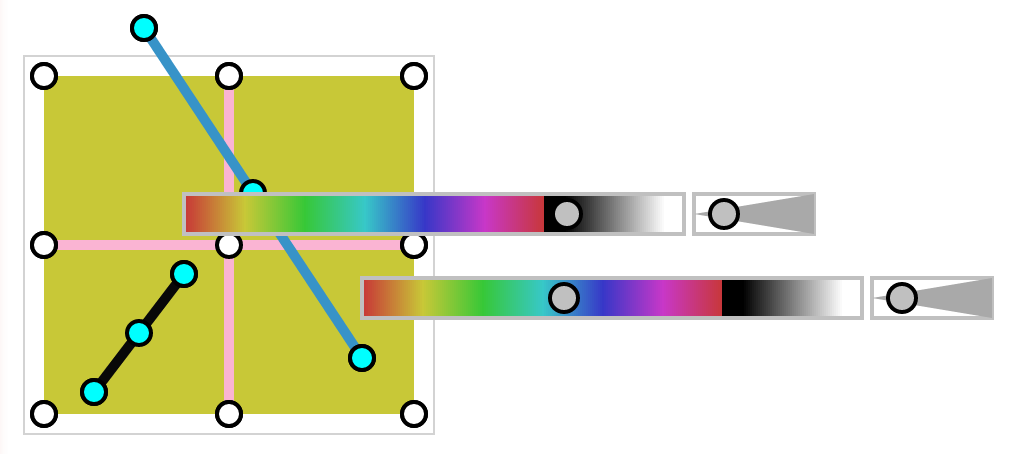}
\end{wrapfigure}

The original version of \sns{}~\cite{sns} allowed the user to
manipulate attributes in the output,
causing the system to immediately infer changes to
constants in the program during the user action.
The size and position of shapes could be adjusted directly.
Attributes without natural visual representations, such
as color and line width, could be manipulated via
\emph{helper widgets} (the sliders in the image). We inherit this functionality
without significant changes.

\subsection{Relating Features}

\begin{figure}[t]
\begin{Verbatim}[commandchars=\\\{\},
                 codes={\catcode`\$=3\catcode`\^=7\catcode`\_=8},
                 numbers=left,numbersep=-6pt,
                 fontsize=\codeSize]
   \verbBoxBlue{(def [rect1\_right rect1\_left] [216 31])}
   \verbBoxBlue{(def [rect1\_bot rect1\_top] [269 100])}
   
   (def rect1
     (let bounds [\verbBoxBlue{rect1\_left rect1\_top rect1\_right rect1\_bot}]
     (let color 60
       [ (rectangle color 'black '0' 0 bounds) ])))
   
   \verbBoxBlue{(def line2\_width 5)}
   \verbBoxBlue{(def line2\_color 202)}
   
   (def line2
     [ (line \verbBoxBlue{line2\_color line2\_width}
             \verbBoxBlue{rect1\_left rect1\_top rect1\_right rect1\_bot}) ])
   
   (def line3
     \verbBoxBlue{(let x2 (* 0.5! (+ rect1\_left rect1\_right))}
     \verbBoxBlue{(let y2 (* 0.5! (+ rect1\_top rect1\_bot))}
       [ (line \verbBoxBlue{line2\_color line2\_width}
               \verbBoxBlue{rect1\_left rect1\_bot} x2 y2) ]
   \verbBoxBlue{))})
 
   (blobs [ rect1 line2 line3 ])
\end{Verbatim}
\vspace{0.05in}

\caption{Overview example after relating features.
The changes compared to \autoref{fig:sns-logo-v1}, made automatically by \sns{},
are highlighted in \verbBoxBlue{bolded blue}.}
\label{fig:sns-logo-v3}
\end{figure}

\begin{figure}[t]

\begin{Verbatim}[commandchars=\\\{\},
                 codes={\catcode`\$=3\catcode`\^=7\catcode`\_=8},
                 numbers=left,numbersep=-6pt,
                 fontsize=\codeSize]
   \verbBoxBlue{(def newGroup4}
   \verbBoxBlue{  ($\lambda$ (line2\_width line2\_color color [left top right bot])}
 
       (def bounds [left top right bot])
     
       (def rect1
         (let bounds [\verbBoxBlue{left top right bot}]
           [ (rectangle color 'black' '0' 0 bounds) ]))
     
       (def line2
         [ (line line2\_color line2\_width
                 \verbBoxBlue{left top right bot}) ])
     
       (def line3
         (let x2 (* 0.5! (+ rect1\_left rect1\_right))
         (let y2 (* 0.5! (+ rect1\_top rect1\_bot))
           [ (line line2\_color line2\_width
                   \verbBoxBlue{left bot} x2 y2) ]
       )))
     
       \verbBoxBlue{[ (group bounds (concat [ rect1 line2 line3 ])) ]))}
 
   (blobs [ \verbBoxBlue{((newGroup4 5 202 60) [31 100 216 269])} ])
\end{Verbatim}
\vspace{0.05in}

\caption{Overview example after grouping and abstracting shapes.
The changes compared to \autoref{fig:sns-logo-v3}, made automatically by \sns{},
are highlighted in \verbBoxBlue{bolded blue}.}
\label{fig:sns-logo-v4}
\end{figure}

Once the basic shapes have been defined,
the next step is to introduce more structure into the program
to define relationships.
To facilitate this process, the Relate phase in \sns{} allows
the user to
(i) select points of interest on the canvas and (ii) declare
that the selected features should be related in some way.

\begin{wrapfigure}{l}{0pt}
\includegraphics[scale=0.28]{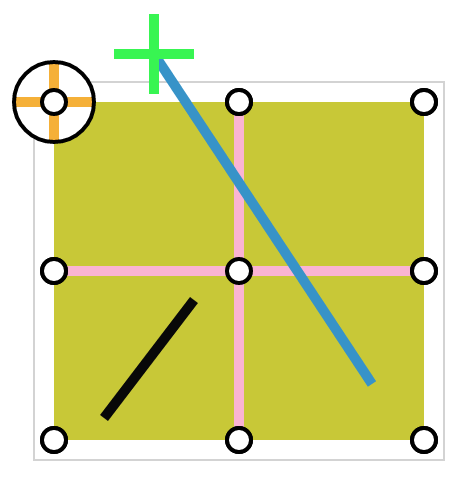}
\end{wrapfigure}
The screenshot on the left shows how features of the initial design can be
selected.
When clicking a positional feature, a crosshair is displayed
that can be used to select position attributes of the point
(see the top-left corner of the rectangle). Selected attributes are
displayed in green (see the top endpoint of the top line).

To start adding relationships, let us make the top-left corner of the
rectangle coincide with the endpoint of the top line (akin to ``snapping''
them together). By clicking the crosshairs of these
two points, we indicate that both x-positions and y-positions,
respectively, ought to be related.
In order to specify our
intended relationship, we click a button labeled Make Equal
which instructs \sns{} to refactor the program
so that, when re-evaluated, it results in the two point values being equal.

We repeat this selection and Make Equal process three
more times, selecting the points at the bottom-left corner,
the bottom-right corner, and the center of the logo.

\begin{wrapfigure}{l}{0pt}
\includegraphics[scale=0.25]{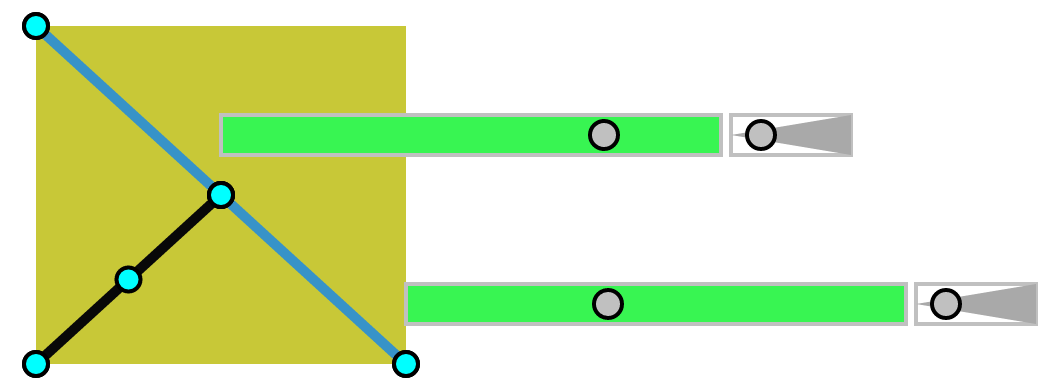}
\end{wrapfigure}
Sliders can also be toggled to relate attributes.
By selecting the two \verb+color+ sliders (displayed in green when
selected) and using Make Equal, and then selecting both line width
sliders and using Make Equal, \sns{} automatically transforms the original
program (\autoref{fig:sns-logo-v1}) to the one shown in \autoref{fig:sns-logo-v3}
so that the width and color of both lines are equal.
We will discuss the transformations in detail in \refSecRelate{}.

\subsection{Grouping Shapes}

Having built a program that implements the logo
in terms of high-level design parameters
(namely, the constants on lines 1, 2, 6, 9, and 10 in
\autoref{fig:sns-logo-v3}),
the final step is to make the design easy to reuse.
To accomplish this, we select the three shapes, click Group to
turn them into a single definition, and then click Abstract to
turn the definition into a reusable function.
Specifically, \sns{} refactors the three selected top-level
definitions in the program
into a single \verb+newGroup4+ definition
and adds function arguments for several design parameters.
The three selected shapes are now generated by a single call to
the \verb+newGroup4+ function.
Thus, the design is ready to be re-used.
We will discuss these transformations
(shown in \autoref{fig:sns-logo-v4})
in more detail in \refSecGroup{}.

To recap, we were able to build the final program entirely using the
new Draw, Relate, and Group tools, resulting in a high-level,
readable function abstracted over the relevant design parameters.
By leveraging the live synchronization inherited from
previous work~\cite{sns}, we can easily change the design
parameters\mdash{}for example, to match the black-and-white
configuration at the beginning of this section\mdash{}by
directly manipulating the logo and sliders.

\section{Tools for Drawing Shapes}
\label{sec:draw}

Having presented an overview of \sns{}, we now discuss
the three new aspects of our work in more detail,
starting with the Drawing tools in this section.
Our choice to use general-purpose
programs as the representation format poses two challenges:
(1) the generated code should be readable so that it can be
extended by the user if desired, and
(2) the code should be structured in a way that facilitates
the subsequent Relate and Group phases.

\begin{wrapfigure}{r}{0pt}
\includegraphics[scale=0.30]{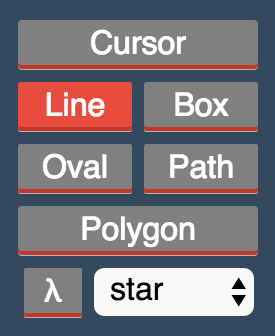}
\end{wrapfigure}
We have defined a library of \emph{stencil programs} for drawing
shapes that satisfy these two goals. In addition, we provide a
way for converting user functions (either manually written or
generated automatically by \sns{} transformations) into stencils
that can be stamped out with direct manipulation.
The screenshot on the right shows the Drawing toolbox in \sns{}.

\levelThree{Program Structure}

A \little{} program \verb+e+ can have arbitrary structure,
but when \verb+e+ is of the form
\verb+(def x1 e1) ... (def xn en) main+, where \verb+main+
is of the form \verb+(blobs [ e1' ... em' ])+, we refer to the
program structure as \emph{simple}.

When drawing a new shape,
simple program structure is preserved by adding the new shape
definition \verb+(def y ey)+ to the end of the definition list
and adding \verb+y+ to the end of the \verb+blobs+ list
(giving it the highest z-order).
When a program \verb+e+ is not simple, we transform the program to
\verb+(let y ey (addShapeToCanvas e y))+, which refers to
a simple library function for list concatenation.

\subsection{Stencils}

We have designed stencil programs for lines, rectangles, ovals,
polygons, and paths (which comprise line and \bez{} curve segments).

\autoref{fig:sns-logo-v1} shows the stencils used to define
rectangles and lines.
The auto-generated code includes randomly chosen colors on lines
4, 9, and 14 for each shape, specified as integers between \verb+0+
and \verb+500+.
The editor draws sliders next to each shape to control these colors,
as well as the line \verb+width+ values on lines 9 and 14, with only
minor differences compared to the previous version of \sns{}~\cite{sns}.

\levelThree{Polygons}

\begin{wrapfigure}{r}{0pt}
\includegraphics[scale=0.28]{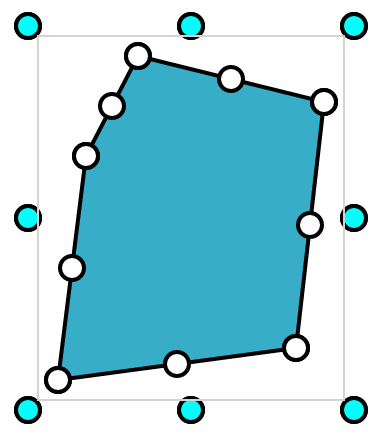}
\end{wrapfigure}
To draw polygons, we use a stencil designed to make it easy
to drag and stretch the entire shape (as one expects from a
direct manipulation interface). For example, for the
shape on the right, \sns{} generates the code below;
the question marks can be ignored for now:

\begin{Verbatim}[commandchars=\\\{\},
                 codes={\catcode`\$=3\catcode`\^=7\catcode`\_=8},
                 fontsize=\codeSize]
  (def polygon1
    (let [left top right bot] [94 101 227 263]
    (let bounds [left top right bot]
    (let [color stroke width] [191 'black' 2]
    (let pcts [[0 1] [0.89? 0.90?] [1 0.14?]
               [0.30? 0] [0.10? 0.31?]]
      [ (stretchyPolygon bounds color stroke width pcts) ]
  )))))
\end{Verbatim}

Notice that \verb+bounds+ defines the bounding box for the polygon,
and each individual point is specified by a pair \verb+[p q]+, where
\verb+p+ (resp. \verb+q+) is the x-position (resp. y-position) of
the point relative to \verb+left+ and \verb+right+
(resp. \verb+top+ and \verb+bot+).
The \little{} library function \verb+stretchyPolygon+
converts the percentages into absolute positions and then draws
a raw SVG \verb+polygon+. The function also draws a
rectangle to surround the polygon.

\begin{wrapfigure}{r}{0pt}
\includegraphics[scale=0.28]{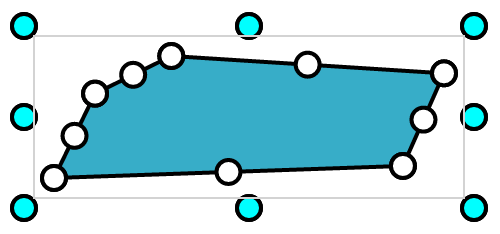}
\end{wrapfigure}
The user can drag the bounding box to stretch all points
accordingly (as shown on the right).
Furthermore, we have designed the \verb+stretchyPolygon+
function so that dragging any of the points will affect the appropriate percentage
values in the program.

For SVG paths, comprising straight line segments, quadratic \bez{} curves,
and cubic \bez{} curves, we have designed and implemented a similar
stencil called \verb+stretchyPath+
that allows an entire path to be stretched easily.
More details can be found in our videos and Web demo.

\subsection{User-Defined Stencils}

The built-in stencils provide basic, scalable shapes.
One of the benefits of using a programming language as the representation
is that user-level designs can be treated on par with built-in primitives.
In particular, a drawable shape may be described as any function
that, given a bounding box argument, draws a shape within the bounding box.
The built-in stencils satisfy this requirement, but so too can user-defined
functions and ones automatically generated by \sns{}.

\begin{wrapfigure}{r}{0pt}
\includegraphics[scale=0.30]{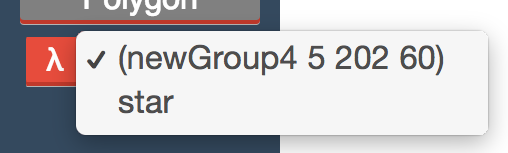}
\end{wrapfigure}
Recall the program in \autoref{fig:sns-logo-v4}, where the
auto-generated \verb+newGroup4+ function comprised the
full logo design. Notice that the last argument to the function
\verb+[left top right bot]+ denotes the bounding box. Our editor
looks for such definitions in the program and adds them to a menu
of user-defined stencils (depicted above).

\begin{wrapfigure}{r}{0pt}
\includegraphics[scale=0.17]{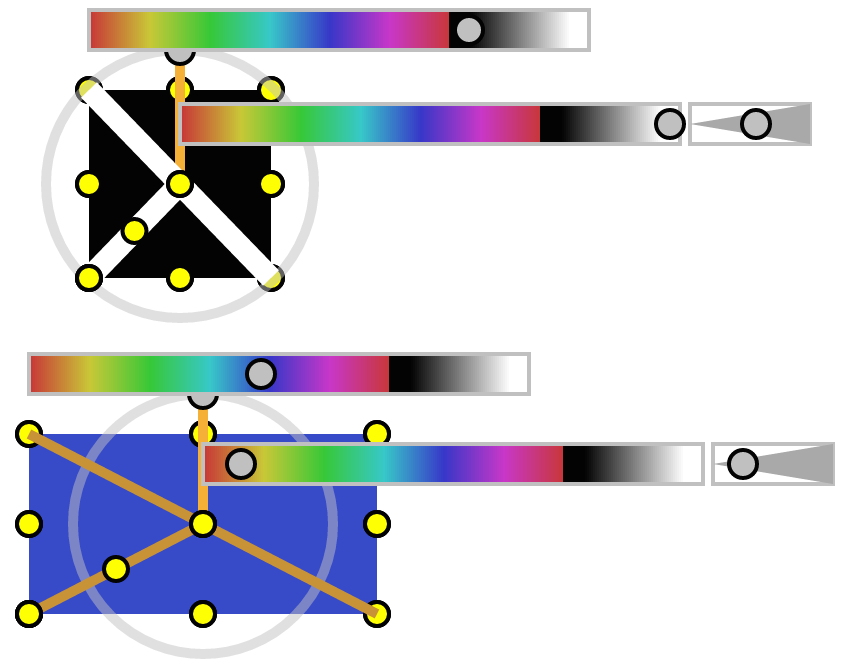}
\end{wrapfigure}
Clicking the Lambda button then allows the user to draw a bounding box
in the canvas, and then \sns{} adds a call to the selected function
(\ie{} lambda) with the bounding box argument.
For example, we may create another instance of the logo using the Lambda
tool, which adds a new function call to \verb+newGroup4+ below.
The arguments to each call serve as the design parameters for that instance.

\begin{Verbatim}[commandchars=\\\{\},
                 codes={\catcode`\$=3\catcode`\^=7\catcode`\_=8},
                 fontsize=\codeSize]
   (blobs [
     ((newGroup4 5 38 232) [39 227 213 317])
     ((newGroup4 11 490 380) [69 55 160 149])
   ])
\end{Verbatim}

\section{Tools for Relating Attributes}
\label{sec:relate}


Once shapes have been drawn, the next step is to encode relationships
in the program.
There are several general concerns if the user were to manually edit
the program:
(1) identifying the relevant portions of the program (\eg{} constants)
that should be related;
(2) boilerplate refactoring needed to move the relevant constants into the
same part of the program;
(3) explicitly programming the intended relationship; and
(4) cleaning up the program after making the above changes.

\sns{} provides a four-step Relate workflow for these concerns.
First, the user selects features of the output (via direct
manipulation) that ought to be related.
Second, \sns{} automatically
refactors the program and \emph{digs} a new \emph{hole expression}
where the selected features can be related in code.
Third, the user or the system \emph{fills} the hole with the desired
relationship.
Finally, \sns{} automatically cleans up unused and unnecessary definitions.


\subsection{Step 1: Select Features}

Each shape is defined by a set of attributes defined by the SVG
specification~\cite{SVG}.
For example, a \verb+rect+ includes \verb+x+, \verb+y+, \verb+width+,
\verb+height+, and \verb+fill+ attribute values,
and a \verb+line+ includes \verb+x1+, \verb+y1+, \verb+x2+, \verb+y2+,
and \verb+stroke+ attribute values.

To select features, the editor displays clickable widgets that are
used to toggle between selection and de-selection.
\begin{wrapfigure}{r}{0pt}
\includegraphics[scale=0.30]{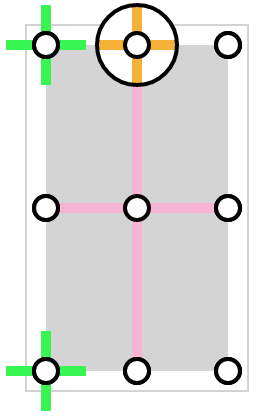}
\end{wrapfigure}
For a positional feature, a crosshair is drawn which can be used to select
the x-position (by clicking the vertical line through the crosshair),
the y-position (the horizontal line), or both (the point in the center
of the crosshair).
For distance attributes like width, height, and radius,
selection lines are drawn that span that distance on the shape.
Furthermore, as described in \refSecOverview{},
a slider can also be toggled (by clicking the backdrop of the slider)
to select the value that it controls.


In addition to the \emph{primitive} SVG features,
the user may want to relate features
that are \emph{derived} in terms of the primitive ones,
such as the bottom-right corner or the center of a rectangle.
For each shape kind, our editor displays selection widgets for
several common derived features. In the screenshot above,
all of the crosshairs except the top-left corner of the rectangle
identify derived features.


\subsection{Step 2: Dig Hole}

\begin{wrapfigure}{r}{0pt}
\includegraphics[scale=0.30]{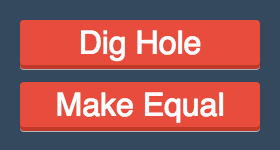}
\end{wrapfigure}
After the user selects features, pressing the Dig Hole button
declares that they ``should be related'' in some way. For example,
let us return to the initial program from the Overview
that contains three unrelated shape definitions (\autoref{fig:sns-logo-v1}).
The first relationship we added was between the top-left corner of
the rectangle and the endpoint of the top line.
After selecting these two points and clicking Dig Hole,
\sns{} refactors part of the program as follows:

\begin{Verbatim}[commandchars=\\\{\},
                 codes={\catcode`\$=3\catcode`\^=7\catcode`\_=8},
                 fontsize=\codeSize]
 ; Lifted Constants
 (def [rect1\_left rect1\_top line2\_x1 line2\_y1] [31 100 81 76])

 ; New Variables and Hole Expression
 (def [rect1\_left' rect1\_top' line2\_x1' line2\_y1']
      {[\verbOrange{rect1\_left  rect1\_top  line2\_x1  line2\_y1} ]})
 (def rect1
   (let [left top right bot] [{rect1\_left' rect1\_top'} 216 269]
     ... ))))

 (def line2
   (let [x1 y1 x2 y2] [{line2\_x1' line2\_y1'} 190 241]
     ... )))
\end{Verbatim}

There are several aspects to notice about the transformed program.
First, the constants that contribute to the selected
features\mdash{}the selected x-values (resp. y-values) are
\verb+31+ and \verb+81+ (resp. \verb+100+ and \verb+76+) on
lines 2 and 8 within the individual definitions\mdash{}have
been lifted into variables in the nearest common scope, in this case,
before the \verb+rect1+ and \verb+line2+ definitions.
The names of new variables
``collect'' names from all of the scopes the constants have
been lifted through. Second, a new variable has been defined for
each lifted constant, suffixed with a prime.
Third, the value of each primed variable is
initialized to its previous value;
we refer to the list expression
that defines these primed variables as the \emph{hole expression}, denoted in orange above.
Finally, the primed variables are used wherever the original constants
had been used before.

After this transformation, the program produces the same output
values because the hole is filled with the previous constant values.
The transformation does, however, create a single place where the
intended relationship can be filled in.


\levelThree{Names for Derived Features}

Primitive features correspond directly to expressions in the program text.
Derived features, however, may not.
If the user wishes to refer to the value of a derived feature
when filling the hole,
she would have to determine how to phrase the value in terms of what
is mentioned in the program.

To simplify this process, we auto-generate named definitions for the
derived features that are selected (in addition to the constant lifting).
Recall that in \refSecOverview{}, we related the endpoint of the bottom line
(a primitive feature defined by the \verb+x2+ and \verb+y2+ values on line 13
of the \verb+line3+ definition in \autoref{fig:sns-logo-v1})
and the center of the rectangle (a derived feature).
To give explicit names for the latter feature, \sns{}
inserts the following:

\begin{Verbatim}[commandchars=\\\{\},
                 codes={\catcode`\$=3\catcode`\^=7\catcode`\_=8},
                 fontsize=\codeSize]
  (def rect1\_boxCX (/ (+ rect1\_left rect1\_right) 2!))
  (def rect1\_boxCY (/ (+ rect1\_top rect1\_bot) 2!))
\end{Verbatim}

The user may then refer to these definitions if they are helpful
for filling in the intended relationship.
The exclamation points on the above constants are
\emph{freeze} annotations that instruct \sns{} not to change those
constants during live synchronization~\cite{sns}.



\subsection{Step 3: Fill Hole}

Once Dig Hole has transformed the program, the next step is
for the user to replace the default initial expression with the
intended relationship. For example, when relating the top-left corner,
we want to use a single constant to control both x-values
and another single constant to control both y-values. Because the
constants themselves do not matter much (they are easy to
change via live synchronization), we might arbitrarily
choose to edit the hole to use only \verb+rect1_left+ and
\verb+rect1_top+ (leaving \verb+line2_x1+ and \verb+line2_x2+
unused):

\begin{Verbatim}[commandchars=\\\{\},
                 codes={\catcode`\$=3\catcode`\^=7\catcode`\_=8},
                 fontsize=\codeSize]
 (def [rect1\_left rect1\_top line2\_x1 line2\_y1] [31 100 81 76])
 (def [rect1\_left' rect1\_top' line2\_x1'  line2\_y1']
      [rect1\_left  rect1\_top  rect1\_left rect1\_top])
\end{Verbatim}

For all of the relationships in the Overview example,
setting constants to be equal is the intended relationship.
In general, however, the user may wish to code an arbitrary
relationship.

\subsection{Step 4: Clean Up}

The Dig and Fill Hole operations introduce several new bindings,
which are helpful during the filling process but may result in
dead or unnecessary code afterwards.
For the filled hole above,
the \verb+line2_x1+ and \verb+line2_y1+ variables are unused, and
the auto-generated \verb+rect1_left'+ and \verb+rect1_top'+
variables are unnecessary since they bind the same values as
\verb+rect1_left+ and \verb+rect1_top+.
Furthermore, some (or most) of the named derived features may not be used.
Therefore, we provide a Clean Up button to perform
inlining, variable renaming, and other transformations to
eliminate such expressions that result from the Select, Dig, and
Fill workflow.

\subsection{Make Equal}

The Relate workflow assists in adding relationships to the
program, but ultimately relies on the user to fill the hole. Ideally, this
step could be automated for common cases.

Because the user often needs to make attributes equal (to ``snap''
positional features together and to make attributes like colors
identical), our implementation provides automated support for filling
the hole to encode equality.
In particular, the Make Equal tool runs Dig Hole,
attempts to fill the hole in a way that makes the selected features equal,
and performs a Clean Up.

To fill the hole, we
(i) consider the program expressions that produced the selected
features, (ii) choose one program constant (\ie{} degree of freedom)
to remove from the program, and (iii) replace that constant with
a variable that is defined in terms of the remaining constants.

\levelThree{Solving Equations}

Our implementation employs and extends a prototype solver to
reason about \emph{value-trace equations} logged when
evaluating a \little{} program~\cite{sns}.
The details about the solver are beyond the scope of this paper.
Below, we describe how it supports two common situations that
arise in our examples.
We write $\ttNumNameVerb{n}{x}$ to denote that the constant literal
$\num{n}$ is bound to the variable \verb+x+ in a program.

One simple case is when relating two features that originate from constants.
For example, to satisfy the equation
$$
\ttNumNameVerb{31}{rect1\_left}=\ttNumNameVerb{81}{line2\_x1}
$$
we arbitrarily choose a constant to eliminate, say,
\verb+line2_x1+, and replace all occurrences of \verb+line2_x1+
with \verb+rect1_left+.

\begin{wrapfigure}{r}{0pt}
\includegraphics[scale=0.28]{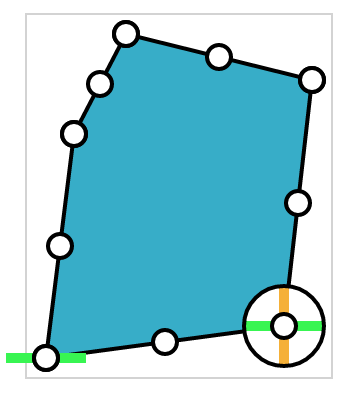}
\end{wrapfigure}
Features are often defined by more complicated expressions than
just constants.
Suppose we want to align the bottom
edge of the stretchy \verb+polygon1+ from before.
Selecting the y-positions of the bottom-left and bottom-right
corners (highlighted in green in the adjacent screenshot)
and pressing Make Equal leads to the following equation:
$$
{\ttNumNameVerb{263}{bot}}
=
{\expAppTwo{\op{+}}
  {\ttNumNameVerb{101}{top}}
  {\expAppTwo{\op{*}}
    {\mathtt{0.90}}
    {\expAppTwo{\op{-}}{\ttNumNameVerb{263}{bot}}{\ttNumNameVerb{101}{top}}}}}
$$
There are three possible degrees of freedom (\ie{} constants) to
remove from the program:
the values of \verb+top+, \verb+bot+, and the percentage \verb+0.90+.
Arbitrarily choosing to remove \verb+top+ or \verb+bot+ would
break the structure of the design.
In this situation, the intent of the original program was that
the same program expression defines all points, and only the
percentages varied. Thus, we would like to only remove the percentage
constants, not the bounding box ones.

\begin{wrapfigure}{r}{0pt}
\includegraphics[scale=0.28]{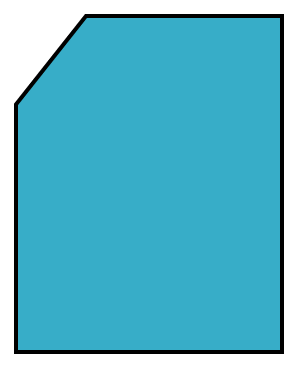}
\end{wrapfigure}
To help disambiguate this common case, a number in the program annotated with
a question mark, written \verb+n?+, serves as a hint to the solver that
it should prefer to remove this constant when given a choice.
Notice that the percentages in \verb+polygon1+ have \verb+?+ annotations
whereas as the \verb+bounds+ constants do not.
We designed the polygon stencil this way so that running Make Equal
on a point inside the bounding box will change only the position of the
point, not the proportions or position of the entire shape.
For the equation above, the constant \verb+0.90+ is removed
from the program and replaced with a variable that binds
\verb+1.00+, the position ratio for the bottom of the bounding box:
\begin{Verbatim}[commandchars=\\\{\},
                 codes={\catcode`\$=3\catcode`\^=7\catcode`\_=8},
                 fontsize=\codeSize]
    (let k3051 1!
    (let pcts [[0 1] [0.89? k3051] [1 0.14?]
               [0.30? 0] [0.10? 0.31?]] ... ))
\end{Verbatim}

Thus, the bottom-right corner is snapped to the bottom of
the bounding box.
Repeating this Make Equal process for each remaining side produces
the ``snip polygon'' shown above.

\section{Tools for Grouping Shapes}
\label{sec:group}

\begin{wrapfigure}{r}{0pt}
\includegraphics[scale=0.30]{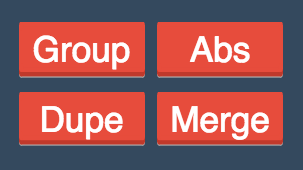}
\end{wrapfigure}
Once the relationships in a program have been structured around a
set of design parameters, the last step is to refactor the program
so that it can be reused easily.
In the Group workflow, \sns{} provides several program transformation
tools for different abstraction patterns, described below.

\subsection{Group}

The Group tool performs three steps.
First, a new top-level
definition is created consisting of the selected blobs
(blobs are either names of top-level definitions or calls to top-level
functions).
Second, a new bounding box \verb+[left top right bot]+
is computed to span the bounding boxes of the selected blobs.
Lastly, so that the entire group scales proportionally
when stretched, the bounding box of each selected blob is
rewritten as a percentage in terms of the new one; this
approach is similar to that used by \verb+stretchyPolygon+.

In \autoref{fig:sns-logo-v4}, notice how the
\verb+newGroup4+ definition contains the three shape definitions
that had previously been at the top-level (in \autoref{fig:sns-logo-v3}).
To rewrite the bounding box of \verb+rect1+ in terms of
the new group bounding box \verb+[left top right bot]+,
the Group transformation would usually generate the following:

\begin{Verbatim}[commandchars=\\\{\},
                 codes={\catcode`\$=3\catcode`\^=7\catcode`\_=8},
                 fontsize=\codeSize]
    (let left  (scaleBetween left right 0)
    (let right (scaleBetween left right 1)
    (let top   (scaleBetween top  bot   0)
    (let bot   (scaleBetween top  bot   1) ... ))))
\end{Verbatim}

But when the relative percentages are equal to \verb+0+ or \verb+1+, \sns{} instead generates
the equivalent but simpler definition on line 7 of \autoref{fig:sns-logo-v4}.
In \refSecEvaluation{}, we will discuss an example where
the hierarchy of bounding boxes is more complicated.

\subsection{Abstract}

The Group tool turns multiple top-level definitions (or calls to top-level
definitions) into a single definition.
Next, the Abstract tool turns a top-level definition into a
function that is abstracted over several parameters.
The challenge for automation is choosing which free variables and constant
literals from the body of the definition to abstract.

To facilitate common use in \sns{}, our heuristic
is to abstract over (non-frozen) constants and, furthermore, those that
have been assigned a name.
Using this approach, \sns{} introduces variables for seven design
parameters on line 2 of \autoref{fig:sns-logo-v4}.

In addition to abstracting such constants, the transformation also
syntactically looks for the bounding box pattern and, when present,
makes a single bounds parameter.
For example, notice how the four bounds parameters
are put into a single list argument and made the last parameter
on line 2 of \autoref{fig:sns-logo-v4}.
With this structure, the Lambda tool recognizes that
it is a new stencil that can be stamped out via the Draw toolbox.
With the new abstraction,
the selected shapes are now generated by a single call to
the \verb+newGroup4+ function (line 23).

\subsection{Duplicate and Merge}

In addition to Abstract, which turns a single blob into a reusable function,
\sns{} supports a second workflow for abstraction.
First, the Duplicate tool copies the code for the selected blob verbatim.
As such, manipulating one copy does not affect the other.

Next, after making changes to some attributes, the user may select
the copies and use the Merge tool. The editor syntactically compares
the definitions for equivalence modulo leaves of the AST (\ie{} constants).
If the definitions are structurally equivalent, the program is transformed to
turn the separate definitions into a single function that is abstracted over
any constants for which not all copies agree.

By using Duplicate and then Merge, the user is, in effect,
explicitly specifying the differences (via the direct manipulation changes)
between the copies, and all other attributes are implicitly encoded to
be equal because they do not become function parameters.
Manipulating any such attribute in one copy will then affect
all copies.
We will discuss examples in \refSecEvaluation{} that use this pattern.

\section{Implementation}
\label{sec:implementation}

\begin{figure}[t]
\begin{center}
\includegraphics[scale=0.37]{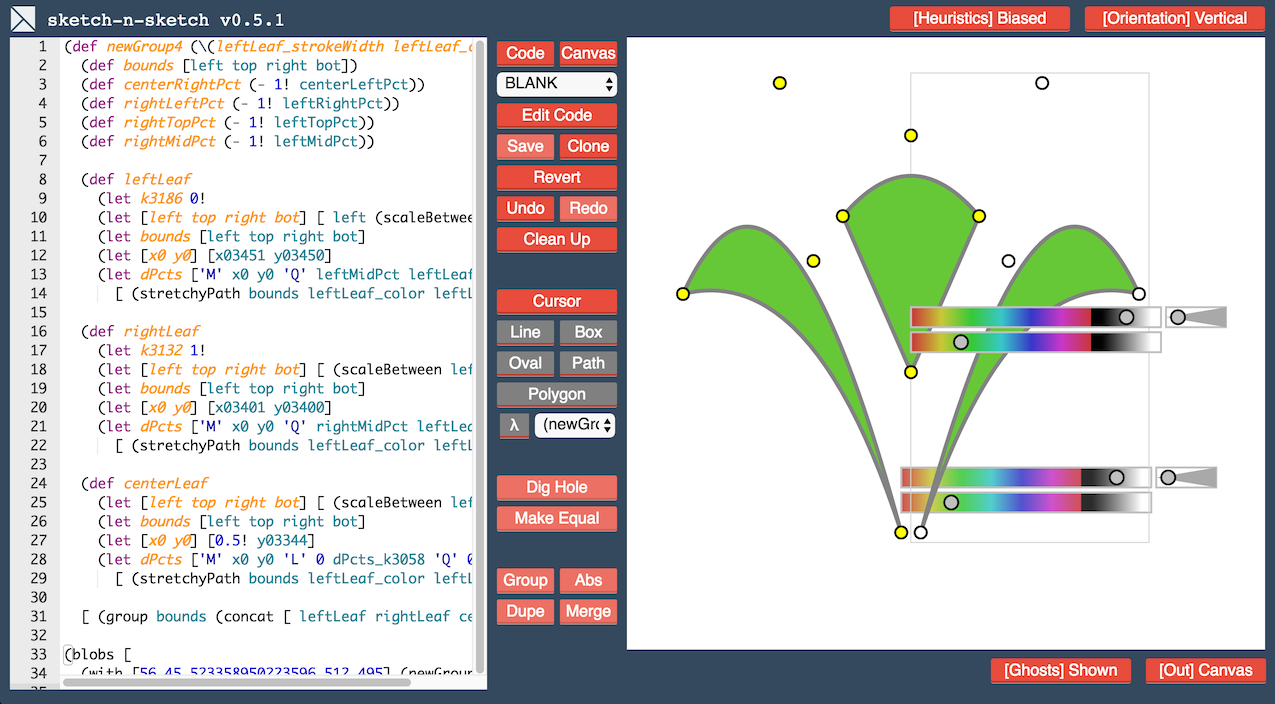}
\end{center}
\caption{Screenshot of \sns{} \textsc{v0.5.1}}
\label{fig:sns-screenshot}
\end{figure}

We have extended \sns{}~\cite{sns}
with tools for the Draw, Relate, and Group workflows.
The extended system is written in more than 13,000 lines of
Elm~\cite{Elm} and JavaScript code (approximately half of
which consists of our extensions) and runs as a standalone
Web application.
\autoref{fig:sns-screenshot} shows a screenshot of our editor,
which displays a program and its output side-by-side with several
tools in between for editing the code and the canvas.

\levelThree{Code Formatting}

Maintaining the readability of the program is one of our design objectives.
Therefore, our implementation takes care to generate and
transform readable code.
The code listings throughout the paper are identical to the ones
generated and transformed by our implementation,
except for minor whitespace changes to improve readability
and to fix a couple of instances where our current implementation
inserts one too many or one too few spaces or lines breaks.

\section{Evaluation}
\label{sec:evaluation}

To evaluate our system, we have implemented three example designs,
in addition to the one described in \refSecOverview{}.
We also discuss examples of how using a general-purpose programming
language offers opportunities for integrating custom user libraries
with direct manipulation tools.
To demonstrate the interactions,
we have recorded several videos and
posted them on the Web.

\subsection{Examples}

In \refSecOverview{}, we described how to build one example
program from scratch. Below, we discuss three more.

\levelThree{\sns{} Logo Revisited}

The logo described in the Overview consisted of two lines drawn over a rectangle. That construction is insufficient if we want to use different colors for the three parts or if we want the negative space to be truly transparent.
Instead, we can construct the logo with three polygons.

\begin{wrapfigure}{l}{0pt}
\includegraphics[scale=0.20]{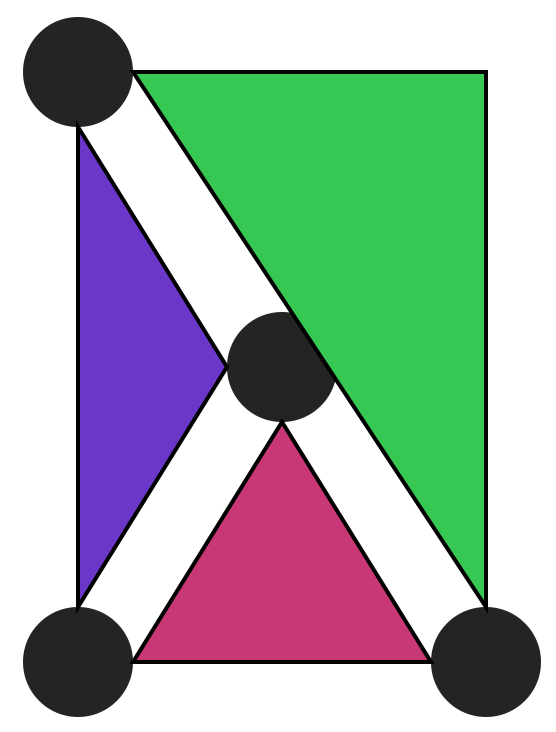}
\end{wrapfigure}
To align and properly space the polygons, we draw four identical helper circles to act as spacers.
We Merge the helper circles to ensure their radii are identical. Then, we use Make Equal to align each corner of each polygon to the edge of an adjacent helper circle.
To make the design rectangular, we align adjacent corners with Make Equal. Similarly, we center the central helper circle by aligning it to a midpoint of each side. The design so far is shown on the left.

To equalize the width and height of the design, we draw a large helper circle (not shown above) and use Make Equal four times to align its bounding box with that of the logo.

\begin{wrapfigure}{l}{0pt}
\includegraphics[scale=0.25]{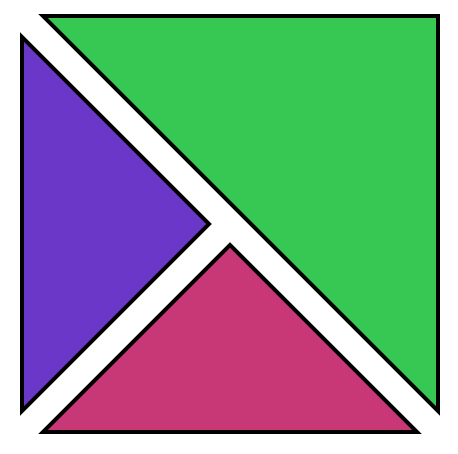}
\end{wrapfigure}
To hide the helper circles from the final output, we could wrap each helper shape in the code with a call to a library function called \verb+ghost+, which would allow us to hide or show them via a button in the interface. Alternatively, since we are done adding relationships to our design, we simply delete the helper shapes. After doing so, we discover that all our constraints are preserved. We can then adjust the spacing between the polygons as well as resize the entire shape using live synchronization.

The entire design is constructed using the drawing tools, 18 uses of
Make Equal, and a single code edit to remove the helper shapes.
However, there is a cost to this ease. The repeated applications of Make Equal
pollute the top level of our code. After four design parameters that
specify the size of the logo and the width of the gap, there
are 12 lines of mathematically correct but unintelligible equations:

\begin{Verbatim}[commandchars=\\\{\},
                 codes={\catcode`\$=3\catcode`\^=7\catcode`\_=8},
                 fontsize=\codeSize]
  (def [polygon6\_top polygon5\_left polygon6\_right]
       [69 88 296])

  (def helper\_r 10.5)

  (def polygon7\_bot (+ (+ (* 0.5! (+ polygon6\_top polygo...
  (def k3105 (/ (- (+ (- polygon6\_right helper\_r) (* 0.5...
  (def polygon7\_top (- (* 0.5! (+ (- polygon7\_bot helper...
  (def [polygon5\_right k3038] [(- (* 0.5! (+ (+ (+ (- po...
  (def k3061 (/ (- (+ polygon5\_right helper\_r) (+ (+ k30...
  (def polygon6\_bot (- (+ (- polygon7\_bot helper\_r) (* 0...
  (def k3063 (/ (- (+ polygon6\_bot helper\_r) polygon7\_to...
  (def polygon5\_top (- polygon6\_top (+ (- 0! (+ helper\_r...
  (def k3103 (/ (- (+ (- polygon5\_top (+ helper\_r helper...
  (def [k3041 polygon5\_bot] [(- polygon7\_top (+ helper\_r...
  (def k3134 (/ (- (+ k3041 helper\_r) polygon5\_top) (- p...
  (def k3141 (/ (- (+ k3038 helper\_r) polygon5\_left) (- ...
\end{Verbatim}

In situations where readability is a priority but where \sns{}
generates expressions like the above, one could fall back on Dig Hole
instead of Make Equal and program the relationships manually.
However, we believe the automatically generated code can be significantly
improved in future work by incorporating a smarter algebraic simplifier;
our current implementation supports only a few local syntactic transformations.

\levelThree{Garden Logo}

\begin{wrapfigure}{r}{0pt}
\includegraphics[scale=0.28]{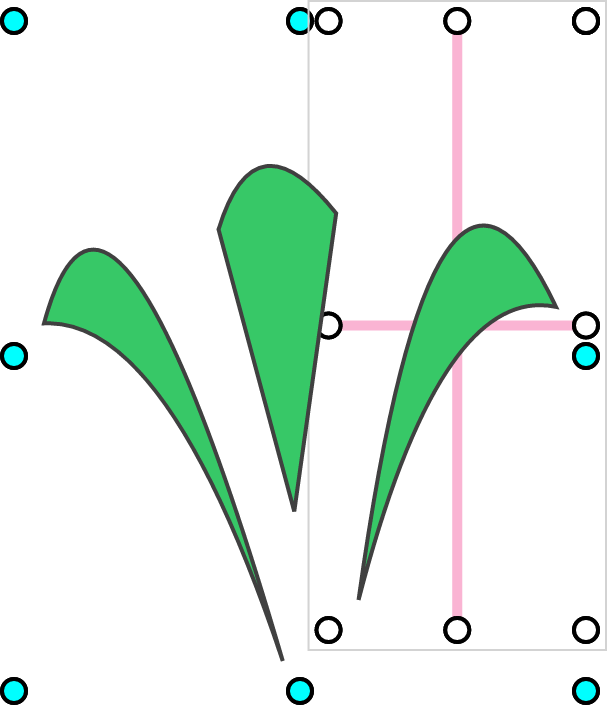}
\end{wrapfigure}
Our next example borrows from the logo of the Chicago Botanic Garden
({\url{www.chicagobotanic.org}}),
which consists of three leaves symmetric across a vertical axis.
We start by drawing three \bez{} curves, and then manually
edit the code to rename the auto-generated names to
\verb+leftLeaf+, \verb+rightLeaf+, and \verb+centerLeaf+.
We equate the \verb+color+, \verb+strokeColor+, and
\verb+strokeWidth+ attributes with several Make Equal operations.

Next, we Group the three leaves, which transforms the program as follows:

\begin{Verbatim}[commandchars=\\\{\},
                 codes={\catcode`\$=3\catcode`\^=7\catcode`\_=8},
                 fontsize=\codeSize]
  (def newGroup
    (def [left top right bot] [55 57 311 362])

    (def leftLeaf ...)
    (def rightLeaf ...)

    (def centerLeaf
      (let left  (scaleBetween left right 0.34?)
      (let top   (scaleBetween top bot 0.10?)
      (let right (scaleBetween left right 0.57?)
      (let bot   (scaleBetween top bot 0.75?) ... )))))

    [ (group bounds
        (concat [ leftLeaf rightLeaf centerLeaf ])) ])
\end{Verbatim}

Notice how the bounding box for the group
comprises the bounding boxes for the three leaves.
As described in \refSecGroup{}, the bounding boxes for the constituent
shapes are rewritten in terms of the new one. For example, the
new bounding box for \verb+centerLeaf+ is specified using percentages;
these can be changed by directly manipulating the box drawn around
\verb+centerLeaf+.

Our final step is to encode horizontal alignment and vertical symmetry.
For horizontal alignment, there are five pairs of y-positions (specified
as percentages) to relate, each of which can be handled by Make Equal.
For vertical symmetry, there are five pairs of x-positions (also specified
as percentages) to relate. For each pair, we use Dig Hole to identify
two percentages \verb+p+ and \verb+q+ and edit the hole to replace
the auto-generated primed variable \verb+q'+
with \verb+(- 1.0! p)+, which encodes symmetry across the central axis.
Lastly, we identify the two percentages in the program that control the
x-positions of the top and bottom of the \verb+centerLeaf+\mdash{}when
hovering over features, \sns{} highlights the relevant
constants in the code\mdash{}and
edit them to be \verb+0.50!+ so that \verb+centerLeaf+ always remains
in the center.
One configuration of the final design is depicted in
\autoref{fig:sns-screenshot}.

Just as Make Equal lets the user ``Align'' shapes
without manual code edits, in future work it would be useful to add
fillers to automatically ``Reflect'' and ``Distribute''
shapes.

\levelThree{Coffee Mug}

\begin{wrapfigure}{r}{0pt}
\includegraphics[scale=0.30]{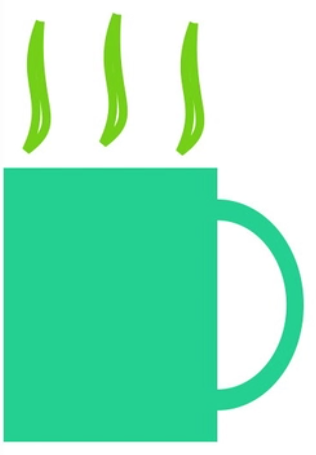}
\end{wrapfigure}
Our last example is a steaming mug of coffee.
We start by drawing two ellipses for the handle and a rectangle
for the body. We select the edges of the inner and outer ellipses,
use Dig Hole to lift their values, and edit the program to center the inner
ellipse inside the outer, but \verb+0.20!+ times smaller.
To affix the handle to the side of the mug, we select the center of the
ellipses and the midpoint of the rectangle's right edge, and align them
with Make Equal.
We also use Make Equal to match the handle color to the body color.
For the steam, we draw one path, use Duplicate to make
two copies, reposition the copies, and use Merge to
re-combine their definitions in the program.

\subsection{UI and Library Co-Design}

The Draw toolbox provides a set of ``built-in'' shapes, but the stencil
code behind them call ordinary \little{}
functions\mdash{}\verb+rectangle+, \verb+line+, \verb+oval+,
\verb+stretchyPolygon+, and \verb+stretchyPath+\mdash{}which
happen to be defined in the standard library.
The Lambda tool is a way to integrate user (or library) customization
with native features.
We describe two additional opportunities for future work below.

\levelThree{Custom Scaling}

The ``stretchy'' semantics provided by \verb+stretchyPolygon+ is
useful in many situations but is not the only scaling semantics the user
may want for a particular shape.
For example, recall the snip polygon shown in \refSecRelate{}.
A different reasonable intention is for the absolute distance between the
snip and the bounding box to remain fixed, even as the polygon is scaled.

Our library provides a \verb+stickyPolygon+ function that provides this
interpretation of scaling, where points ``stick'' to the nearest corner.
To draw the same polygon described by \verb+polygon1+ in \refSecDraw{},
the following code uses absolute offsets, rather than percentages, to describe
the points; each pair \verb+[[x dx] [y dy]]+ identifies the offset
\verb+[dx dy]+ from a particular corner \verb+[x y]+:

\begin{Verbatim}[commandchars=\\\{\},
                 codes={\catcode`\$=3\catcode`\^=7\catcode`\_=8},
                 fontsize=\codeSize]
  (let [left top right bot] [94 101 227 263]
  (let offsets
    [ [[left 2?] [bot 0]] [[right -6?] [bot -21?]]
      [[right 0] [top 30?]] [[left 37?] [top 0]]
      [[left 0] [top 49?]] ]
  [ (stickyPolygon bounds color stroke width offsets) ]
\end{Verbatim}

An even different scaling behavior commonly found in existing tools combines
both of the above; features, such as rounded or snipped corners, may
stretch up to a certain point after which they stick at a fixed distance.
Because \little{} is a general-purpose language, we can write a
\verb+stretchySnipPolygon+ function to encode this behavior as a small
variation on the previous functions.

\levelThree{Custom Features}

For each kind of shape, \sns{} draws selection widgets for
a set of derived features.
These features are currently hard-coded in our implementation,
but there might be other features relevant for a particular design.
\begin{wrapfigure}{r}{0pt}
\includegraphics[scale=0.20]{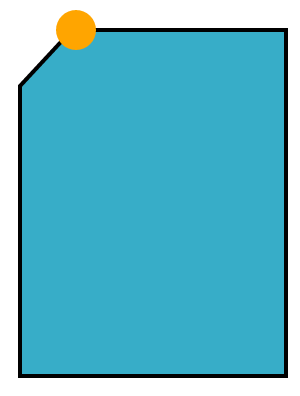}
\end{wrapfigure}
In the adjacent screenshot, we modified our program to draw a helper
dot to identify the maximum snip distance of a \verb+stretchySnipPolygon+.
It may be useful to allow users and libraries
to customize the derived features displayed and manipulated.

\section{Discussion}

To wrap up, we compare \sns{} to related work,
augmenting the landscape described in the Introduction.
We also discuss limitations of our approach and opportunities
for future work.

\subsection{Combining Programming and Direct Manipulation}

Programmatic and direct manipulation have been combined in a myriad of diverse
configurations. Prior systems have generally emphasized direct manipulation over
programming, with the underlying programs expressed in
domain-specific languages or with special data structures.

\levelThree{Scriptable Direct Manipulation Editors}
Some direct manipulation systems
provide scripting APIs that allow users to run
editor commands programmatically. To ease the programming burden, some editors
allow the user to record their actions into a macro script
(\eg{} SolidWorks~\cite{SolidWorksMacro})
or echo equivalent scripting commands for every action
to facilitate copy-paste scripting (\eg{} Maya~\cite{Maya}).

\levelThree{Parametric Computer-Aided Design}
In \emph{parametric feature-based CAD} systems,
user actions are concatenated into a procedure (\ie{} script), hidden from
the user, that specifies the
step-by-step creation of the design. Features of new elements can be defined
in terms of the parameters of existing elements. When the user changes
a parameter, all commands in the script that depend on it are automatically
re-run\mdash{}for example, if the user resizes the main cylinder of a
screw, the screw head defined to be 1.5x wider will automatically be resized as
well~\cite{BPECAD}. Grouping and simple repetition are also generally
supported, and the EBP system~\cite{EBPCAD} additionally implemented programming
by demonstration interactions to specify loops and conditionals.

\levelThree{Procedural Modeling}
In \emph{procedural modeling} systems,
specialized algorithms automatically generate complex 2D or 3D content
such as trees, terrain, or road networks. Some of these systems allow algorithmic
parameters to be adjusted by directly manipulating the output.
However, the algorithms themselves are usually built-in to the system and cannot be altered by the user~\cite{PMSurvey}.

Algorithms based on generative shape grammars generally allow text-based grammar modifications. Richer direct manipulation interactions have also been
explored
to change individual rule parameters~\cite{KrecklauKobbeltGrammars,LippVisualGrammars},
to copy-paste rules~\cite{ProceduralCopyPaste},
to manipulate grammars visually~\cite{NodeBasedGrammar,PatowGrammar,LippVisualGrammars},
and to infer grammars from sample output~\cite{InversePM1,InversePM2,InversePM3}.

\levelThree{Programming by Example or Demonstration}
Many PBE~\cite{PBE} and PBD~\cite{PBD} systems for graphics editing support
operations analogous to the drawing, relating, and grouping of this paper
but use domain-specific representations for the underlying program
(\eg{}~\cite{Chimera,Metamouse,Mondrian}).
Two recent systems are QuickDraw, which infers both constraints~\cite{QuickDraw2012}
and procedural repetition~\cite{QuickDraw2014} based on a user-provided sketch,
and Drawing Dynamic
Visualizations~\cite{VictorDrawing}, in which direct manipulation operations
construct a program as a graphical history that supports constraints, looping,
and parameterized abstraction.

\levelThree{Our Approach}

In contrast, we start with a general-purpose language and codify
the user's actions by transforming the program. These two features\mdash{}a general-purpose language and code
transformations\mdash{}are the key differentiators of our approach;
all of the above systems either rely upon a domain-specific language or
provide only basic features for editing the code.

\subsection{Live Synchronization}

Our previous version of \sns{} required that the initial program be written using
traditional text-based editing, a burden reduced by our new techniques.

Once written, \emph{live synchronization}~\cite{sns} allowed changes to
the output of a program to immediately change appropriate constants
in the program.
To do so, \sns{} records run time-traces that
relate the input program to its output; these relationships
form \emph{value-trace equations}.
To synthesize program updates, the system attempts
to solve one value-trace equation per updated attribute
by changing the value of exactly one constant in the program.
When there are ambiguities (\ie{} multiple constants to change),
\sns{} uses heuristics to automatically choose without
asking the user for help.

When extending \sns{}, we inherited the live synchronization
approach\mdash{}along with the limitations that stem from the one-equation,
one-constant design\mdash{}with only minor changes.
One modification was to add a new \verb+BOX+ primitive specified by
\verb+left+, \verb+right+, \verb+top+, and \verb+bot+ values
which translates to the SVG \verb+rect+ primitive.

There are two ways in which our new features are designed with
the behavior of live synchronization in mind.
First, we carefully designed our scalable polygon and path
stencils so that they interact well with live synchronization;
directly manipulating the exterior points also manipulates the
bounding box, and directly manipulating the interior
points manipulates the appropriate constants when running
\sns{} in ``biased'' heuristics mode~\cite{sns}.

Second, the ways that our Group operations transform the program
favor an approach where bounding boxes are defined with
constants in the program and inner shapes are defined relative
to the bounding box. Such programs lend themselves to more
intuitive live synchronization than ones where shape attributes
are defined directly and bounding boxes are implicitly derived
from them.

Nevertheless, the ideas behind our shape stencils and program
transformations could be re-purposed to favor other programming
patterns instead.

\subsection{Constraint-Oriented Programming}

Constraint-oriented programming systems, including
SketchPad~\cite{SketchpadThesis} and ThingLab~\cite{Borning1981}
among others~\cite{Sketchpad14,ChecksAndBalances,Juno},
allow users to specify declarative relationships that augment
procedural programs.
Constraint solvers (\eg{}~\cite{cassowary}) attempt to satisfy
the declared relationships.

Although our Relate workflow suggests the declarative feel of constraints, our
representation format is a concrete,
deterministic program. Since we do not rely on additional constraint
or solver state, our functional programs may be more easily reused
in other, general-purpose programming domains.

Several user interaction techniques have been explored for defining and breaking
constraints~\cite{PBM,BriarConstraintDrawing,ChimeraSnapshots}, some of which
might be applied to our system to improve our UI for declaring relationships.

\subsection{Automated Refactoring}

Some of our program transformations\mdash{}Clean Up, Group, Abstract,
and Merge\mdash{}can be viewed as variations on common program refactorings,
whereas some\mdash{}Dig Hole\mdash{}do not have direct analogs.

Our Dig Hole approach lifts all relevant constants in
the program into a common scope. An alternative design might look
for additional expressions (beyond just constants) where an
intended relationship might be filled in.

Currently, our Group transformations work only for programs in
the \emph{simple} structure described earlier.
For cases where the program has more complicated
structure, and for abstracting finer-grained pieces of code than
just top-level definitions, it would be useful to develop a
more general methodology for syntactic abstraction of source
programs, probably taking into account run-time traces~\cite{sns}.
Such generalized approaches might benefit from interactive,
visual editors such as those described below.

\subsection{Structured and Visual Editors}

It may be useful in future work to design a
more visual, structured editor~\cite{Teitelbaum1981,Gandalf}\mdash{}particularly
one like Barista~\cite{Barista} that also
supports unstructured, text-based editing\mdash{}specifically
for the workflow that arises in \sns{}.

For example, because our tools automatically generate and transform
code, the user may often want to easily rename and reorder definitions.
In addition, whereas our current Merge and Abstract transformations
use specific design heuristics to decide what parts of
a program expression to abstract, it would be useful to provide
a way for the user to interact with the system during the process.
Finally, although the named definitions for derived features
inserted by Dig Hole are useful, it would be better to provide
a visual connection between the rendered features and the generated
program expressions.

In general, such improvements may be lead to new direct manipulation
features for code editing itself. Widgets~\cite{sns} and so-called
``scrubbing tools''~\cite{VictorDrawing,McDirmid} can be viewed as
examples of this notion.

\subsection{Conclusion}

The new drawing, relating, and grouping interactions we added to
\sns{} are potentially useful enhancements for users that currently
choose languages (\eg{} Processing) and libraries (\eg{} D3) to programmatically
generate 2D vector graphics.

In the longer-term, we view this
work as a milestone towards a vision where
(a) experts may choose to use tools
that mix programming and direct manipulation in a variety of
domains, and where (b) novices get some of the benefits of
programming by (i) having expert library writers customize tools
with new purely GUI-based features, (ii) learning a bit of
programming through the live connection and semi-automated
programming tools, and (iii) applying and extending
programming-by-example to these settings. In pursuit of this vision, we will
continue to explore how rich direct manipulation capabilities can provide
interactive and intuitive environments for general-purpose
programming languages.

%
%
%
%
%
\balance

\newpage

\bibliographystyle{acm-sigchi}


\end{document}